\documentclass[12pt,aip]{article}
\begin{document}
\title{Lyapunov stability of flowing MHD plasmas surrounded by resistive
walls}
\author{H. Tasso\footnote{het@ipp.mpg.de}, G. N.
Throumoulopoulos\footnote{gthroum@uoi.gr} \\
$^\star$
Max-Planck-Institut f\"{u}r Plasmaphysik \\
Euratom Association \\  85748 Garching bei M\"{u}nchen, Germany \\
$^\dag$ University of Ioannina,\\ Association Euratom-Hellenic
Republic,\\ Department of Physics, GR 451 10 Ioannina, Greece}
\maketitle
\newpage
\begin{abstract}
A general stability condition for plasma-vacuum systems with
resistive walls is derived by using the Frieman Rotenberg lagrangian
stability formulation [Rev. Mod. Phys. 32, 898 (1960)]. It is shown
that the Lyapunov stability limit for external modes does not depend
upon the gyroscopic term but upon the sign of the perturbed
potential energy only. In the absence of dissipation in the plasma
such as viscosity, it is expected that the flow cannot stabilize the
system.
\end{abstract}


\newpage

\section{Introduction}

In a previous theorem \cite{pt} a general instability statement was
proved concerning static plasmas surrounded by resistive walls. This
statement is related to the problem of resistive wall modes (RWM) in
the magnetohydrodynamic (MHD)literature. In particular, the impact
of plasma rotation on the mode has been investigated theoretically
in
 Refs.\cite{BoWa}-\cite{HaGi} and experimentally
 in Refs.\cite{ReGa}-\cite{BeSaBe}.
 See also Ref.\cite{gpt} for mode
calculations in a cylindrical case and the recent review paper of
Ref.\cite{ChOk}.  In the present paper we show that the general
Lyapunov analysis can be extended to stationary plasmas surrounded
by resistive walls even if their resistivity is time-dependent as
investigated in Ref.\cite{tt}.

In section II we discuss the Lagrangian stability investigation of
Ref.\cite{fr} in the context of free boundary displacements through
an interface between plasma and vacuum. The Lyapunov functional is
constructed in section III while conclusions and discussions are
left to section IV.

\section{Stability equation}

The linearized Lagrangian displacement vector ${\mbox{\boldmath
$\xi$}}$ about an ideal MHD equilibrium with flow obeys the
equations derived in Ref.\cite{fr}:

\begin{equation}
\rho_{0}  \ddot{\mbox{\boldmath $\xi$}} + 2\rho_{0}{\bf
v_{0}}\cdot\nabla\dot{\mbox{\boldmath $\xi$}} + {\bf
F}\mbox{\boldmath$\xi$} = 0,
\end{equation}
where $\rho_{0}$ is the equilibrium mass density and ${\bf v_{0}}$
the flow velocity. The antisymmetric $2\rho_{0}{\bf
v_{0}}\cdot\nabla$ operator is sometime called the "gyroscopic"
operator and ${\bf F}$ is given by
\begin{equation}
{\bf F}\mbox{\boldmath$\xi$} = \nabla(\gamma
p_{0}\nabla\cdot{\mbox{\boldmath $\xi$}} + {\mbox{\boldmath
$\xi$}}\cdot\nabla p_{0} - {\bf B}\cdot {\bf Q}) + {\bf
B}\cdot\nabla{\bf Q} + {\bf Q}\cdot\nabla{\bf B} +
\nabla\cdot(\rho_{0}{\mbox{\boldmath $\xi$}}{\bf
v_{0}}\cdot\nabla{\bf v_{0}} - \rho_{0}{\bf v_{0}}{\bf
v_{0}}\cdot\nabla{\mbox{\boldmath $\xi$}}),
\end{equation}
where $p_{0}$ is the equilibrium pressure and ${\bf B}$ the
equilibrium magnetic field. ${\bf Q}$ =
$\nabla\times({\mbox{\boldmath $\xi$}}\times{\bf B})$ is the
perturbed magnetic field. The operator ${\bf F}$ of Eq.(2) contains
in particular flow terms so generalizing the corresponding operator
of the static case \cite{bf}.

After fixing the gauge by annihilating  the scalar potential, the
vector potential ${\bf A}$ in the region outside the plasma is
governed by Ohm's law
\begin{equation}
\nabla\times\nabla\times {\bf A} = - \sigma \dot {\bf A}
\end{equation}
where $\sigma $ is the space and time dependent conductivity in the
outer region.$\sigma $ vanishes for a true vacuum.

At the interface between the plasma and the outer region, we must
fulfil the condition
\begin{equation}
{\bf n}\times {\bf A} = - ({\bf n}\cdot {\mbox{\boldmath $\xi$}})
{\bf B_{0}}
\end{equation}
where ${\bf B_{0}}$ is the equilibrium magnetic field at the vacuum
side of the interface whose normal vector is ${\bf n}$. ${\bf A}$
has to vanish at infinity or be normal to superconducting walls, if
there are any. Equations (1) to (4) reduce to the equations (1) to
(3) of Ref.\cite{tt} if the flow vanishes.

The gyroscopic operator $2\rho_{0}{\bf v_{0}}\cdot\nabla$ is
antisymmetric and the scalar product $2\int \dot{\mbox{\boldmath
$\xi$}}\cdot( \rho_{0}{\bf v_{0}}\cdot\nabla)\dot{\mbox{\boldmath
$\xi$}} d\tau$ as a volume integral vanishes for ${\bf
n}\cdot{\mbox{\boldmath $\xi$}} = 0$ at the plasma boundary. If
${\bf n}\cdot{\mbox{\boldmath $\xi$}} \neq 0$ at the interface the
scalar product vanishes also if the mass density vanishes at the
boundary. To make sure that this is true we use
\begin{equation}
\nabla\times({\mbox{\boldmath $\xi$}}\times\rho_{0}{\bf v_{0}}) =
\rho_{0}{\bf v_{0}}\cdot\nabla{\mbox{\boldmath $\xi$}} -
{\mbox{\boldmath $\xi$}}\cdot\nabla (\rho_{0}{\bf v_{0}}) -
\rho_{0}{\bf v_{0}}\nabla\cdot{\mbox{\boldmath $\xi$}},
\end{equation}
\begin{equation}
\nabla({\mbox{\boldmath $\xi$}}\cdot\rho_{0}{\bf v_{0}}) =
\rho_{0}{\bf v_{0}}\cdot\nabla{\mbox{\boldmath $\xi$}} +
\rho_{0}{\bf v_{0}}\times\nabla\times{\mbox{\boldmath $\xi$}} +
{\mbox{\boldmath $\xi$}}\cdot \nabla\rho_{0}{\bf v_{0}} +
{\mbox{\boldmath $\xi$}}\times\nabla\times\rho_{0}{\bf v_{0}}.
\end{equation}
Note that the term ${\mbox{\boldmath $\xi$}} \nabla\cdot(\rho_0 {\bf
v_{0}})$
 has  vanishing contribution to Eq.(5) because of the equilibrium
 continuity equation.
Adding equations (5) and (6) we obtain
\begin{equation}
\nabla({\mbox{\boldmath $\xi$}}\cdot\rho_{0}{\bf v_{0}}) =
2\rho_{0}{\bf v_{0}}\cdot \nabla{\mbox{\boldmath $\xi$}} -
\rho_{0}{\bf v_{0}}\nabla\cdot{\mbox{\boldmath
$\xi$}} - \nabla\times({\mbox{\boldmath $\xi$}}\times\rho_{0}{\bf v_{0}}) +\\
\rho_{0}{\bf v_{0}}\times\nabla\times{\mbox{\boldmath $\xi$}} +
{\mbox{\boldmath $\xi$}}\times\nabla\times\rho_{0}{\bf v_{0}}.
\end{equation}
After taking the scalar product of Eq.(7) with ${\mbox{\boldmath
$\eta$}}$ and using
\begin{equation}
\nabla\cdot[{\mbox{\boldmath $\eta$}}\times({\mbox{\boldmath
$\xi$}}\times\rho_{0}{\bf v_{0}})] = ({\mbox{\boldmath
$\xi$}}\times\rho_{0}{\bf v_{0}})\cdot\nabla\times{\mbox{\boldmath
$\eta$}} - {\mbox{\boldmath
$\eta$}}\cdot\nabla\times({\mbox{\boldmath $\xi$}}\times\rho_{0}{\bf
v_{0}}),
\end{equation}
\begin{equation}
\nabla\cdot[({\mbox{\boldmath $\eta$}}\cdot \rho_{0}{\bf
v_{0}}){\mbox{\boldmath $\xi$}}] = ({\mbox{\boldmath $\eta$}}\cdot
\rho_{0}{\bf v_{0}})\nabla\cdot{\mbox{\boldmath $\xi$}} +
{\mbox{\boldmath $\xi$}}\cdot \nabla ({\mbox{\boldmath
$\eta$}}\cdot{\rho_{0}\bf v_{0}}),
\end{equation}
we integrate over the plasma volume and transform the volume
integrals of the divergences to surface integrals at the interface
to obtain
\begin{equation}
\int_{V}{\mbox{\boldmath $\eta$}}\cdot(\rho_{0}{\bf
v_{0}}\cdot\nabla){\mbox{\boldmath $\xi$}} d\tau = -
\int_{V}{\mbox{\boldmath $\xi$}}\cdot ({\rho_{0} \bf
v_{0}}\cdot\nabla){\mbox{\boldmath $\eta$}}  d \tau
\end{equation}
if either ${\bf n}\cdot{\mbox{\boldmath $\xi$}}$ = ${\bf
n}\cdot{\mbox{\boldmath $\eta$}}$ = $0$ for internal modes or
$\rho_{0}{\bf v_{0}}$ = $0$ at the interface for external modes.

\section{Lyapunov functional}

Taking the scalar product of Eq.(1) with $\dot{{\mbox{\boldmath
$\xi$}}}$ and Eq.(3) with $\dot{\bf A}$ and integrating over the
plasma volume and the vacuum region respectively we obtain
\begin{equation}
\frac{d(K + W)}{dt} = - (\sigma \dot{{\bf A}}, \dot{{\bf A}})
\end{equation}
where
\begin{eqnarray*}
 W = \frac{ (\mbox{\boldmath $\xi$} , {\bf F}\mbox{\boldmath $\xi$})_{p} +
{\bf (A, \nabla \times \nabla \times A )}_{or}}{2}, \\ K =
\frac{(\rho_{0}{\bf \dot{\mbox{\boldmath $\xi$}}}, {\bf
\dot{\mbox{\boldmath $\xi$}}})_{p}}{2},
\end{eqnarray*}
the subscripts $p$ and $or$ being related to the plasma and the
outer region respectively. The parentheses denote the scalar
products
\begin{eqnarray*}
 (\mbox{\boldmath $\eta$} ,\mbox{\boldmath $\mu$})_{p} =
\int_{p} {\bf (\mbox{\boldmath $\eta$}
,\mbox{\boldmath $\mu$})} d \tau, \\
 (\mbox{\boldmath $\eta$} ,\mbox{\boldmath $\mu$})_{or} =
\int_{or} {\bf (\mbox{\boldmath $\eta$} ,\mbox{\boldmath
$\mbox{\boldmath $\mu$}$})} \tau.
\end{eqnarray*}

 Since $\sigma$ is positive, the
quantity $K + W$ has a negative time derivative for external modes
according to Eq.(11). This allows us to apply the method of
Lyapunov, which states that if $K + W$ has no definite sign, then
the system is "Lyapunov unstable". For internal modes with ${\bf A}
= 0$ the condition is only sufficient. The Hermitian form $W$ could
be minimized in the same way as the $\delta W$ of the energy
principle \cite{bf} for static MHD equilibria, but no numerical code
for minimizing $W$ has been developed yet. For a straight tokamak
with homogeneous axial flow and without walls $W$ is indefinite if
it is indefinite in the static case. This follows from the fact that
the stability property does not depend upon the inertial frame
chosen. Therefore, according to the above Lyapunov study the
introduction of resistive walls cannot stabilize the plasma-wall
system with finite constant plasma velocity. One should note,
however, that Lyapunov stability is equivalent to spectral stability
if all modes of the spectrum are considered. This is not in
agreement of the result of Ref.\cite{SmJa}. In general, the entire
spectrum of the plasma-wall system is not as easy to investigate as
the determination of the sign of $W$.

Eq.(11) reduces to Eq.(4) of Ref.\cite{pt,tt} if the velocity terms
in $F$ vanish. It extends the Lyapunov functional introduced in
Ref.\cite{pt,tt} to the present general flow case. This means that
if $W$ is positive definite the system is stable, but if it is
indefinite the system is unstable. So the gyroscopic operator does
not affect stability of external modes but growth rates, eigenvalues
and eigenfunctions only.

\section{Discussion and conclusions}

The question of stability of external modes for toroidal ideal
plasmas with flows surrounded by resistive walls is decided by the
sign of the symmetric potential operator only, which is a drastic
simplification so that numerical investigation of stability becomes
only slightly more difficult than in the static case for which many
codes have been developed.

Moreover the Lyapunov stability property holds also for internal
modes if a dissipation operator is added to the gyroscopic operator
in (1) since this addition would not change the sign of the right
hand side of (11). However, if we add a viscosity to the physical
system under investigation then (1) would be modified by such an
operator acting on $\dot{{\mbox{\boldmath$\xi$}}}$ but also by
another operator acting on ${\mbox{\boldmath$\xi$}}$ which
represents the "circulatory" forces as considered in Ref.\cite{ta}.
The operator on $\dot{{\mbox{\boldmath$\xi$}}}$ is proportional to
the viscosity itself but the "circulatory" operator goes with
viscosity times the velocity ${\bf v_{0}}$. Only the latter operator
can modify the stability property, so a minimum velocity is needed
for this operator to be able to stabilize the system. However, the
quantitative investigation of such situations is difficult as
explained in Ref.\cite{ta} and Ref.\cite{tt1}. We should also
mention that the plasma viscosity is a tensor known only in the
collisional regime while a convincing derivation for the weak
collisionality regime of fusion plasmas has not been done yet.
Therefore, it may turn out that the stability of hot plasmas must be
treated by kinetic equations like Vlasov or Fokker-Planck for which
the present stability procedure does not apply.

Finally, if for physical interpretations of the experiment,
eigenmodes and growth rates of (1) to (4) (and possibly nonlinear
effects) are needed, then the investigation of such situations will
be much more difficult especially for toroidal plasmas. Numerical
techniques for the search of complex eigenvalues must be developed
in the first place. (See e.g. \cite{ke}).

\newpage

\begin{center}

{\large\bf Acknowledgements}

\end{center}

Part of this work was conducted during a visit of one of the authors
(G.N.T.) to the Max-Planck-Institut f\"{u}r Plasmaphysik, Garching.
The hospitality of that Institute is greatly appreciated. The
present work was performed under the Contract of Association ERB
5005 CT 99 0100 between the European Atomic Energy Community and the
Hellenic Republic.The views and opinions expressed herein do not
necessarily reflect those of the European Commission.

\end{document}